\documentclass{article}
\def\lb{\label}
\def\be{\begin{equation}}
\def\ee{\end{equation}}
\def\ba{\begin{eqnarray}}
\def\ea{\end{eqnarray}}

\def\ol{\overline}
\def\bb{\bibitem}

\def\e{{\rm e}}
\begin{document}
\title{
   \begin{flushright} \begin{small}
     LAPTH-1087/05, DTP-MSU/05-02,   \\
  \end{small} \end{flushright}
\vspace{.5cm}
The Letelier-Gal'tsov spacetime revisited}
\author
{G. Cl\'ement$^{a}$\thanks{email: gclement@lapp.in2p3.fr} , D.V.
Gal'tsov$^{b}$\thanks{email: galtsov@mail.phys.msu.ru} and P.S.
Letelier$^{c}$\thanks{email: letelier@ime.unicamp.br}
\\ \\ {\small $^{a}$Laboratoire de
Physique Th\'eorique LAPTH (CNRS), B.P.110, F-74941
Annecy-le-Vieux cedex, France} \\
{\small $^{b}$Department of Theoretical
Physics, Moscow State University, 119899, Moscow, Russia} \\
{\small $^{c}$Departamento de Matem\'atica
Aplicada-IMECC, Universidade
Estadual de Campinas, 13083-970 Campinas, S.P., Brazil}
}

\maketitle

\begin{abstract}
Contrary to a recent claim by Anderson [{\em The Mathematical Theory
of Cosmic Strings}, I.O.P.Publishing, Bristol 2003], we show that the
Letelier-Gal'tsov metric does represent a system of crossed straight
infinite cosmic strings moving with arbitrary constant relative velocities.
\end{abstract}

\bigskip

In a recent book on ``the mathematical theory of cosmic strings"
\cite{and}, Anderson devotes a section to ``the proper status of the
Letelier-Gal'tsov `crossed-string' metric". This metric describes in a
single coordinate patch the Gott \cite{gott} geometry  that consists in the
gluing of two separate spacetimes, each one representing a tilted
moving string. After describing the
construction of \cite{letgal}, Anderson argues that ``the
Letelier-Gal'tsov metric seems capable of describing an arbitrary
number of open strings with essentially arbitrary shapes and in
arbitrary states of relative motion", which according to him
``suggests something that was perhaps evident from the start, namely
that [this metric] is just the static parallel string metric written
in an obscure coordinate system''. He then proceeds to substantiate
this claim by showing that the geodesic distance between two strings
is actually constant. However, as we shall show here, his proof fails
due to the omission of an integration constant in the approximate
evaluation of a certain integral near a given string. Correcting some
misprints contained in the original paper \cite{letgal}, we shall
also show that the Letelier-Gal'tsov metric indeed represents a
system of crossed straight cosmic strings moving with arbitrary but
constant relative velocities.
\\

The Letelier-Gal'tsov ``multiple moving crossed cosmic strings
spacetime'' is obtained from the Minkowski spacetime
\be\lb{min}
ds^2 = dt^2 - dz^2 - dZd\ol{Z}
\ee
by the singular coordinate transformation defined by the line integral
\be\lb{Z}
Z(\zeta) = X + iY = \int_{\zeta_0}^{\zeta} \prod_{i=1}^N
(\xi-\alpha_i(t,z))^{-4\lambda_i}d\xi
\ee
($\zeta = x+iy$) with $\zeta_0$ fixed. This line element can be rewritten as
\be\lb{mmccs}
ds^2 = dt^2 - dz^2 - \e^{-4V}(d\zeta + Fdt + Gdz) (d\ol\zeta +
\ol{F}dt + \ol{G}dz)
\ee
with
\ba
V & = & \sum_{i=1}^N\lambda_i\ln|\zeta-\alpha_i|^2\,,\\ F & = &
\prod_{i=1}^N(\zeta-\alpha_i)^{4\lambda_i} \int_{\zeta_0}^{\zeta}
\frac{d\xi}{\prod_{k=1}^N
(\xi-\alpha_k)^{4\lambda_k}}\sum_{j=1}^N\frac{4\lambda_j\dot{\alpha}_j}
{\xi-\alpha_j}\,, \\ G & = &
\prod_{i=1}^N(\zeta-\alpha_i)^{4\lambda_i} \int_{\zeta_0}^{\zeta}
\frac{d\xi}{\prod_{k=1}^N
(\xi-\alpha_k)^{4\lambda_k}}\sum_{j=1}^N\frac{4\lambda_j\alpha'_j}
{\xi-\alpha_j}\,.
\ea
\\

Consider for definiteness two moving strings of equal tension located
at $\alpha_{\pm} = \pm \alpha(t,z)$. Then,
\be\lb{Z2}
Z(\zeta) = \int_{\zeta_0}^{\zeta} (\xi^2-\alpha^2)^{-4\lambda}d\xi,
\ee
and
\be\lb{F}
F(\zeta) =
8\lambda\alpha\dot{\alpha}(\zeta^2-\alpha^2)^{4\lambda}I(\zeta)\,,
\qquad G(\zeta) =
8\lambda\alpha{\alpha}'(\zeta^2-\alpha^2)^{4\lambda}I(\zeta)\,,
\ee
with
\be\lb{I}
I(\zeta) = \int_{\zeta_0}^{\zeta} (\xi^2-\alpha^2)^{-4\lambda-1}d\xi.
\ee
Near $\zeta = +\alpha$, the derivative of $I$ behaves as
\be
\frac{\partial I}{\partial\zeta} = (2\alpha)^{-4\lambda-1}
(\zeta-\alpha)^{-4\lambda-1} - (1+4\lambda)(2\alpha)^{-4\lambda-2}
(\zeta-\alpha)^{-4\lambda} + \cdots,
\ee
which integrates to
\be
I = C_{\lambda}(\alpha,\zeta_0) -
\frac1{4\lambda}(2\alpha)^{-4\lambda-1} (\zeta-\alpha)^{-4\lambda} -
\frac{1+4\lambda}{1-4\lambda}
(2\alpha)^{-4\lambda-2}(\zeta-\alpha)^{1-4\lambda} + \cdots
\ee
There is no reason for the integration constant $C_{\lambda}
(\alpha,\zeta_0)$, which measures the ``finite part'' of $I(\zeta)$,
to vanish. It follows that
\be\lb{Fs}
F(\zeta) = -\dot{\alpha} + 4\lambda C_{\lambda}(\alpha,\zeta_0)
(2\alpha)^{1+4\lambda} \dot{\alpha}(\zeta-\alpha)^{4\lambda}-
\frac{4\lambda}{1-4\lambda} \frac{\dot{\alpha}}{\alpha}(\zeta-\alpha)
+ \cdots\,,
\ee
and a similar expansion for $G(\zeta)$. For $\lambda < 1/4$, the
second term of (\ref{Fs}) dominates the third (the corresponding term
is absent from the first Eq. (14) of \cite{letgal}).
\\

At fixed $z$ and $t$, the geodesic distance between the two strings
is $|Z(\alpha)-Z(-\alpha)|$. From the above, $Z(\alpha)$ varies
according to
\be\lb{dZ}
dZ = (\zeta^2-\alpha^2)^{-4\lambda}(d\zeta + F\,dt + G\,dz)|_{\zeta =
\alpha}
  \simeq 8\lambda C_{\lambda}(\alpha,\zeta_0)\alpha(\dot{\alpha}\,dt
+ \alpha'\,dz)\,, \ee showing that the strings are indeed tilted
and moving. Note that Anderson, following exactly the same line of
reasoning, obtained $dZ \simeq
O(\zeta-\alpha)^{1-4\lambda}(\dot{\alpha}\,dt + \alpha'\,dz) \to
0$, because in his evaluation of $F(\zeta)$ near the string he
omitted the integration constant $C_{\lambda}$ and thus included
only the first and third terms of (\ref{Fs}).
\\

One can directly evaluate $Z(\alpha)$  by computing
$dZ(q\alpha)/d\alpha)$ ($q$ constant, $0<q<1$) and taking the limit
$q\to1$. Choosing for convenience $\zeta_0 = 0$ (``center-of-mass''),
\be\lb{Zq}
Z_q(\alpha) \equiv Z(q\alpha) = \int_{0}^{q\alpha}
(\xi^2-\alpha^2)^{-4\lambda}d\xi.
\ee
The derivative is
\be
\frac{dZ_q}{d\alpha} = q(q^2\alpha^2-\alpha^2)^{-4\lambda} +
8\lambda\alpha \int_{0}^{q\alpha} (\xi^2-\alpha^2)^{-4\lambda-1}d\xi.
\ee
Also, computing (\ref{Zq}) by integrating by parts, one obtains
\be
(1-8\lambda)Z_q = q\alpha(q^2\alpha^2-\alpha^2)^{-4\lambda} +
8\lambda\alpha^2 \int_{0}^{q\alpha}
(\xi^2-\alpha^2)^{-4\lambda-1}d\xi = \alpha\frac{dZ_q}{d\alpha}.
\ee
This differential equation is integrated by
\be\lb{ZK}
Z_q(\alpha) = K_{\lambda}(q) \alpha^{1-8\lambda},
\ee
with $K_{\lambda}(q)$ an integration constant.
This makes sense for $q=1$ (direct computation gives $K_0(1) = 1$,
$K_{1/8}(1) = -i\pi/2$), and thus gives the relative motion of the
cosmic strings for all $\alpha(t,z)$. Comparing with (\ref{dZ}), one
obtains the relation between the integration constants $C_{\lambda}$
and $K_{\lambda}$
\be
C_{\lambda}(\alpha,0)
=\frac{1-8\lambda}{8\lambda}K_{\lambda}(1)\,\alpha^{-1-8\lambda}.
\ee
This dependence of $C_{\lambda}(\alpha,0)$ on $\alpha$ can also be directly
obtained from the definition (\ref{I}) of $I$ by a scaling argument.
\\

Also, the relative string motion is, contrary to appearance,
not arbitrary. In the special case of
parallel strings, the system can be trivially reduced (by omitting
the $dz^2$ in (\ref{mmccs})) to a (2+1)-dimensional system of conical
singularities. As conical singularities correspond to point
particles, one must require for consistency \cite{dj} that these
follow geodesics of the spacetime (this approach was used
succesfully in \cite{BH,sig} to study the dynamics of systems
of black holes and
conical singularities). More generally, it has been shown
that the world-sheets of self-gravitating cosmic strings are totally
geodesic submanifolds \cite{vickers, UHIM}. In the present case, the
spacetime is flat outside the cosmic strings, and the solution of the
geodesic equations for the $Z_i \equiv Z(\zeta = \alpha_i)$ is
straightforwardly given in terms of initial data $Z_{0i}$, $v_i$ and
$w_i$ by
\be\lb{geo}
Z_i(t,z) = Z_{0i} + v_it + w_iz.
\ee
In the case of our example of two strings in the center-of-mass
frame, comparison of (\ref{ZK}) and (\ref{geo}) leads to the solution
\be\lb{geo2}
\alpha(t,z) = (\beta t + \gamma z + \delta)^{1/(1-8\lambda)},
\ee
with $\beta$, $\gamma$ and $\delta$ (initial ``velocities'',
``inclinations" and ``positions'') complex integration constants
(proportional to $v$, $w$, and $Z_0$).
\\

The relative motion of two cosmic strings has been reduced, via
(\ref{min}), (\ref{Z2}) and (\ref{geo2}), to the (1+1)-dimensional
free motion of a point particle. The world-line of this particle may
always be Lorentz-transformed to the time axis in the case of a
subluminal velocity ($v^2 < w^2$), or to the $z$-axis in the case of
a superluminal velocity ($v^2 > w^2$). Correspondingly, it follows
that the two-cosmic string system may be Lorentz-transformed to a
static system of two crossed strings in the first case, or to a
system of two moving parallel strings in the second case, in
accordance with the results of \cite{GGL}.
\\

Thus the metric (\ref{mmccs}) indeed describes a system of
infinite straight strings arbitrarily oriented and moving with
respect to each other. The strings are gravitationally
interacting, so one could expect gravitational radiation to occur.
But this space-time does not contain gravitational waves and does
not exhibit radiative features. In the case of parallel strings,
this can be easily understood as a consequence of the absence of
gravitons in $2+1$ gravity. For two crossed strings undergoing a
superluminal (subluminal) collision this follows from the
possibility of finding a Lorentz frame in which the strings are moving
parallel (static). In the limiting case of a ``luminal''
collision the absence of gravitational radiation follows from the
existence of the above exact solution.
\\

It is worth noting that this simplicity is lost once other string
interactions are taken into account. Two crossed strings moving
with a superluminal relative velocity {\em do} radiate a massless
second rank antisymmetric tensor field (string axion) if the
strings are charged with respect to this field. This effect was
recently considered using perturbation theory \cite{GMK}.
Therefore a system of strings interacting with a two-form field
can hardly be described by a simple solution like (\ref{mmccs}).

\end{document}